# Single-Symbol-Decodable Differential Space-Time Modulation Based on QO-STBC

Chau Yuen
cyuen@i2r.a-star.edu.sg
Institute for Infocomm Research,
Singapore

Yong Liang Guan
eylguan@ntu.edu.sg
Nanyang Technological University,
Singapore

Tjeng Thiang Tjhung
tjhungtt@i2r.a-star.edu.sg
Institute for Infocomm Research,
Singapore

**Abstract** – We present a novel differential space-time modulation (DSTM) scheme that is single-symbol decodable and can provide full transmit diversity. It is the first known single-symbol-decodable DSTM scheme not based on Orthogonal STBC (O-STBC), and it is constructed based on the recently proposed Minimum-Decoding-Complexity Quasi-Orthogonal Space-Time Block Code (MDC-QOSTBC). We derive the code design criteria and present systematic methodology to find the solution sets. The proposed DSTM scheme can provide higher code rate than DSTM schemes based on O-STBC. Its decoding complexity is also considerably lower than DSTM schemes based on Sp(2) and double-symbol-decodable QO-STBC, with negligible or slight trade-off in decoding error probability performance.

## I. INTRODUCTION

Differential Space-Time Modulation (DSTM) schemes are capable of providing transmit diversity without the need for channel estimation. First-generation DSTM schemes were designed based on unitary matrices [1-6]. In [4], a unitary DSTM scheme based on Orthogonal STBC (O-STBC), whose decoding can be achieved by simple linear processing, is proposed. But its maximum achievable code rate is only ¾ for four transmit antennas and ½ for eight transmit antennas. In [5,6], new DSTMs to achieve higher code rate than the O-STBC DSTM were proposed. The DSTM in [5] is designed based on Quasi Orthogonal STBC (QO-STBC), while the DSTM in [6] is based on a new non-linear STBC. Both these schemes are double-symbol-decodable. Due to the unitary requirement on their basis space-time codes, the DSTMs mentioned above employ phase-shift keying (PSK) constellation, which becomes less power efficient as the constellation size increase. To alleviate this limitation, DSTMs based on quasi-unitary matrices from O-STBC [7-9] and double-symbol-decodable QO-STBC with non-constant matrix norm [10] were subsequently proposed.

In this paper, we propose a new *quasi-unitary* DSTM scheme which is *single-symbol decodable* based on the Minimum-Decoding-Complexity QO-STBC (MDC-QOSTBC) described in [11, 12]. We will derive its code design criteria, design its required constellation set, and compare its error probability performance and decoding complexity with existing DSTM schemes.

## II. Review of DSTM Scheme Based on Quasi-Unitary Matrices

### *A. Signal Model*

Consider a MIMO channel with $N_T$ transmit antennas and $N_R$ receive antennas. Let $\mathbf{H}_t$ be the $N_R \times N_T$ channel gain matrix at a time $t$. Thus the $ik^{\text{th}}$ element of $\mathbf{H}_t$ is the channel coefficient for the signal path from the $k^{\text{th}}$ transmit antenna to the $i^{\text{th}}$ receive antenna. Let $\mathbf{X}_t$ be the $N_T \times N_T$ square codeword transmitted at a time $t$. Then, the received $N_R \times N_T$ signal matrix $\mathbf{R}_t$ can be written as

$$\mathbf{R}_t = \mathbf{H}_t \mathbf{X}_t + \mathbf{N}_t \tag{1}$$

where $\mathbf{N}_t$ is the additive white Gaussian noise.

At the start of the transmission, we transmit a known unitary codeword $\mathbf{X}_0$. The codeword $\mathbf{X}_t$ transmitted at a time $t$ is then differentially encoded by

$$\mathbf{X}_t = a_{t-1}^{-1} \mathbf{X}_{t-1} \mathbf{U}_t \tag{2}$$

where $\mathbf{U}_t$ is a quasi-unitary square matrix (i.e. $\mathbf{U}_t\mathbf{U}_t^{\mathrm{H}} = a_t^2\mathbf{I}$) called the *code matrix* that contains information of the transmitted data and $a_{t-1}^2$ is the diagonal element of $\mathbf{U}_{t-1}\mathbf{U}_{t-1}^{\mathrm{H}}$. Note that the value of $a_t$ may change with time since $\mathbf{U}_t$ is quasi-unitary. However, to ensure that the total average transmission power is maintained constant, it is required that

$$\mathrm{E}\left(\mathbf{X}_t\mathbf{X}_t^{\mathrm{H}}\right) = \mathrm{E}\left(\mathbf{U}_t\mathbf{U}_t^{\mathrm{H}}\right) = \mathbf{I}, \text{ or } \mathrm{E}\left(a_t^2\right) = 1. \tag{3}$$

If we assume that the channel remains unchanged during two consecutive code periods, i.e. $\mathbf{H}_t = \mathbf{H}_{t\text{-}1}$, then the received signal $\mathbf{R}_t$ at time $t$ can be expressed as

$$\mathbf{R}_t = a_{t-1}^{-1}\mathbf{R}_{t-1}\mathbf{U}_t + \tilde{\mathbf{N}}_t \tag{4}$$

where $\tilde{\mathbf{N}}_t = -a_{t-1}^{-1}\mathbf{N}_{t-1}\mathbf{U}_t + \mathbf{N}_t$ (5)

Since $\tilde{\mathbf{N}}_t$ is not independent of $a_{t\text{-}1}$ and $\mathbf{U}_t$, the optimal DSTM decoder has high computational complexity, hence a near-optimal differential decoder has been proposed in [8] to estimate $\mathbf{U}_t$ as follows:

$$\begin{aligned}
\hat{\mathbf{U}}_t &= \arg\min_{\mathbf{U}_t\in\mathcal{U}} \left\|\mathbf{R}_t - a_{t-1}^{-1}\mathbf{R}_{t-1}\mathbf{U}_t\right\|^2 \\
&= \arg\min_{\mathbf{U}_t\in\mathcal{U}} \mathrm{tr}\left(a_{t-1}^{-2}\mathbf{U}_t^{\mathrm{H}}\mathbf{R}_{t-1}^{\mathrm{H}}\mathbf{R}_{t-1}\mathbf{U}_t - a_{t-1}^{-1}\mathbf{R}_t^{\mathrm{H}}\mathbf{R}_{t-1}\mathbf{U}_t - a_{t-1}^{-1}\mathbf{U}_t^{\mathrm{H}}\mathbf{R}_{t-1}^{\mathrm{H}}\mathbf{R}_t\right) \\
&= \arg\min_{\mathbf{U}_t\in\mathcal{U}} \mathrm{tr}\left[a_{t-1}^{-2}\mathbf{R}_{t-1}^{\mathrm{H}}\mathbf{R}_{t-1}\mathbf{U}_t\mathbf{U}_t^{\mathrm{H}} - 2a_{t-1}^{-1}\,\mathrm{Re}\left(\mathbf{R}_t^{\mathrm{H}}\mathbf{R}_{t-1}\mathbf{U}_t\right)\right]
\end{aligned} \tag{6}$$

where $\mathcal{U}$ denotes the set of all possible code matrices, tr[.] represents the trace of a matrix, $\|.\|^2$ represents the Frobenius norm and $a_{t\text{-}1}$ can be estimated from the previous decision $\hat{\mathbf{U}}_{t-1}$ [1]. It has been shown in [8] that the performance loss of the near-optimal differential decoder in (6) compared to the optimal differential decoder is small, but the near-optimal differential decoder could lead to a significant reduction in decoding complexity. Hence in this paper, we adopt the near-optimal decoder for our proposed DSTM scheme.

*B. Design Criteria*

An essential difference between quasi-unitary DSTM and unitary DSTM is that $a_{t-1}$ in (4), (5) and (6) is a not a constant value in quasi-unitary DSTM. So the optimal design of quasi-unitary DSTM should take $a_{t-1}$ into account, as it causes variation in the noise variance. However, the optimal design criteria for quasi-unitary DSTM are not known as $a_{t-1}$ itself depends on the previous codeword, hence the code design problem becomes complicated. In order to keep the code design problem tractable, we invoke the averaging condition of (3) that $\mathrm{E}\left(a_t^2\right)=1$, i.e. $a_{t-1}$ is constant on the average. In this way, the code design problem of quasi-unitary DSTM becomes similar to that of unitary DSTM, and we can now apply its established code design criteria (*Rank and Determinant* criteria) on quasi-unitary DSTM. Though approximate, this design approach will be shown to give good solutions that achieve single-symbol-decodability and full diversity while maintaining good decoding performance.

We now summarize the design criteria for unitary DSTM. The decoding error probability performance of unitary DSTM scheme has been analyzed in [2]. Specifically, the diversity level that can be achieved is:

$$\mathrm{Min}\left[\mathrm{rank}\left(\mathbf{U}_k-\mathbf{U}_l\right)\right] \qquad \forall k \neq l . \tag{7}$$

To achieve full transmit diversity, the minimum rank in (7) must be equal to $N_T$. For a full-rank DSTM code, its *coding gain* is defined in [2][4] as

$$\mathrm{Min}\left[N_T \times \det\left(\left(\mathbf{U}_k-\mathbf{U}_l\right)\left(\mathbf{U}_k-\mathbf{U}_l\right)^{\mathrm{H}}\right)^{1/N_T}\right] \qquad \forall k \neq l . \tag{8}$$

To achieve optimum decoding error probability, the coding gain in (8) must be maximized.

---

[1] This may lead to error propagation, but it has been shown in [8-10] that the effect of error propagation is small. This will also be confirmed by our simulation results in Figure 4 later in this paper.

## III. New DSTM Scheme Based on MDC-QOSTBC

### *A. Minimum-Decoding-Complexity QO-STBC (MDC-QOSTBC)*

MDC-QOSTBC was recently proposed in [11, 12] as a class of single-symbol decodable QO-STBC whose maximum likelihood decoding only needs to jointly decode two real symbols. Hence its decoding complexity is second only to O-STBC. To construct an MDC-QOSTBC, which consists of $K$ sets of dispersion matrices denoted as $\{\mathbf{A}, \mathbf{B}\}$ with code length $T$ for $N_T$ transmit antennas ($N_T$ an even number greater than two), from an O-STBC which consists of $K/2$ sets of dispersion matrices denoted as $\{\underline{\mathbf{A}}, \underline{\mathbf{B}}\}$ with code length $T / 2$ for $N_T / 2$ transmit antennas, the following dispersion matrix mapping rules (described in [11, 12]) can be used:

$$\text{Rule \#1: } \mathbf{A}_i = \frac{1}{\sqrt{K}}\begin{bmatrix} \underline{\mathbf{A}}_i & \mathbf{0} \\ \mathbf{0} & \underline{\mathbf{A}}_i \end{bmatrix} \qquad \text{Rule \#2: } \mathbf{B}_i = \frac{1}{\sqrt{K}}\begin{bmatrix} \mathbf{0} & j\underline{\mathbf{A}}_i \\ j\underline{\mathbf{A}}_i & \mathbf{0} \end{bmatrix}$$
$$\text{Rule \#3: } \mathbf{A}_{\left(\frac{K}{2}\right)+i} = \frac{1}{\sqrt{K}}\begin{bmatrix} j\underline{\mathbf{B}}_i & \mathbf{0} \\ \mathbf{0} & j\underline{\mathbf{B}}_i \end{bmatrix} \qquad \text{Rule \#4: } \mathbf{B}_{\left(\frac{K}{2}\right)+i} = \frac{1}{\sqrt{K}}\begin{bmatrix} \mathbf{0} & \underline{\mathbf{B}}_i \\ \underline{\mathbf{B}}_i & \mathbf{0} \end{bmatrix} \qquad \text{for } 1 \le i \le \frac{K}{2} \quad (9)$$

where the factor $1/\sqrt{K}$ is for normalizing the transmission power. For example, a rate-1 MDC-QOSTBC [11, 12], $\mathbf{C}$, with $N_T = T = K = 4$ is shown in (10).

$$\mathbf{C} = \sum_{i=1}^{K} c_i^{\mathrm{R}}\mathbf{A}_i + jc_i^{\mathrm{I}}\mathbf{B}_i = \frac{1}{\sqrt{4}}\begin{bmatrix} c_1^{\mathrm{R}} + jc_3^{\mathrm{R}} & -c_2^{\mathrm{R}} + jc_4^{\mathrm{R}} & -c_1^{\mathrm{I}} + jc_3^{\mathrm{I}} & c_2^{\mathrm{I}} + jc_4^{\mathrm{I}} \\ c_2^{\mathrm{R}} + jc_4^{\mathrm{R}} & c_1^{\mathrm{R}} - jc_3^{\mathrm{R}} & -c_2^{\mathrm{I}} + jc_4^{\mathrm{I}} & -c_1^{\mathrm{I}} - jc_3^{\mathrm{I}} \\ -c_1^{\mathrm{I}} + jc_3^{\mathrm{I}} & c_2^{\mathrm{I}} + jc_4^{\mathrm{I}} & c_1^{\mathrm{R}} + jc_3^{\mathrm{R}} & -c_2^{\mathrm{R}} + jc_4^{\mathrm{R}} \\ -c_2^{\mathrm{I}} + jc_4^{\mathrm{I}} & -c_1^{\mathrm{I}} - jc_3^{\mathrm{I}} & c_2^{\mathrm{R}} + jc_4^{\mathrm{R}} & c_1^{\mathrm{R}} - jc_3^{\mathrm{R}} \end{bmatrix} \qquad (10)$$

where the symbols $c_i$ represent the transmitted symbols and the superscripts $^{\mathrm{R}}$ and $^{\mathrm{I}}$ represent the real and imaginary parts of a transmitted symbol respectively.

For the MDC-QOSTBC constructed by (9), it has been shown in [13] that,

$$\mathbf{CC}^{\mathrm{H}} = \frac{\alpha}{K}\begin{bmatrix} \mathbf{I}_{N_T/2} & \mathbf{0} \\ \mathbf{0} & \mathbf{I}_{N_T/2} \end{bmatrix} + \frac{\beta}{K}\begin{bmatrix} \mathbf{0} & \mathbf{I}_{N_T/2} \\ \mathbf{I}_{N_T/2} & \mathbf{0} \end{bmatrix} \qquad (11)$$

where
$$\alpha = \sum_{i=1}^{K} |c_i|^2$$
$$\beta = 2\sum_{i=1}^{K/2} -c_i^{\mathrm{R}}c_i^{\mathrm{I}} + c_{(K/2)+i}^{\mathrm{R}}c_{(K/2)+i}^{\mathrm{I}} \qquad (12)$$

and the minimum determinant of its codeword distance matrix is:

$$\det_{\min} = \frac{1}{K}\left[\left(\Delta_i^{\mathrm{R}}\right)^2 - \left(\Delta_i^{\mathrm{I}}\right)^2\right]^{N_T} \tag{13}$$

where $\Delta_i$, $1 \le i \le 4$, represents the possible error in the $i^{\text{th}}$ transmitted constellation symbol. That is, $\Delta_i = c_i - e_i$ if the receiver decides erroneously in favor of $e_i$ when actually $c_i$ has been transmitted. In order to achieve full diversity and optimum coding gain in accordance with the design criteria defined in Section IIB, the value of (13) has to be non-zero and maximized.

*B. New DSTM Based on MDC-QOSTBC*

To use the MDC-QOSTBC codeword **C** in (10) as the code matrix $\mathbf{U}_t$ of a quasi-unitary DSTM scheme, **C** must have the property that $\mathbf{CC}^{\mathrm{H}} = \gamma\mathbf{I}$ where $\gamma$ is a constant value. Based on (11) and (12), this requires that $\beta = 0$ and $\mathrm{E}(\alpha) = K$, where E(.) denotes the mean value. If the code symbols $c_i$ in **C** are memoryless PSK or QAM symbols, then $\beta$ in (12) may not be zero and $\mathbf{CC}^{\mathrm{H}} \neq \gamma\mathbf{I}$. Hence **C** with memoryless PSK or QAM constellation is not quasi-unitary and cannot be used as a DSTM code matrix.

To achieve $\beta = 0$ for **C**, we know from (12) that $c_i^{\mathrm{R}} c_i^{\mathrm{I}}$ must be equal to $c_{K/2+i}^{\mathrm{R}} c_{K/2+i}^{\mathrm{I}}$. For example, if $K = 4$, we must have $c_1^{\mathrm{R}} c_1^{\mathrm{I}} = c_3^{\mathrm{R}} c_3^{\mathrm{I}}$ and $c_2^{\mathrm{R}} c_2^{\mathrm{I}} = c_4^{\mathrm{R}} c_4^{\mathrm{I}}$. This can be achieved by mapping $c_i$ to a constellation symbol $z_k = x_k + jy_k$ such that $x_k y_k = v \ \forall k$ where $v$ is a constant. Next, to obtain $\mathrm{E}(\alpha) = K$, Equation (12) suggests that $c_i$ must be mapped to the constellation symbol $z_k = x_k + jy_k$ with the additional constraint of $\mathrm{E}\left(x_k^2 + y_k^2\right) = 1$, assuming that all transmitted symbols have the same power. Finally, to maximize the coding gain, the constellation symbols $z_k$ should also be designed to achieve maximum possible value of $\det_{\min}$ in (13). In summary, the collection of constellation symbols $z_k = x_k + jy_k$, $1 \le k \le M$, denoted as constellation set $\mathcal{M}$, for use in a quasi-unitary DSTM based on MDC-QOSTBC, should be designed with the following requirements:

(i) Quasi-Unitary Criterion: $x_k y_k = v$

(ii) Power Criterion: $\mathrm{E}\left(x_k^2 + y_k^2\right) = 1$ (14)

(iii) Performance Criterion: maximize $\mathrm{Min}\left\{\left[\left(\Delta x_{kl}\right)^2 - \left(\Delta y_{kl}\right)^2\right]^{N_T}\right\}$ where $v$ can be any constant value, and $\Delta x_{kl} = x_k - x_l$, $\Delta y_{kl} = y_k - y_l$ for $1 \le k \ne l \le M$.

The spectral efficiency of the resultant DSTM scheme constructed using a MDC-QOSTBC with code rate $R$ will be ($R$ $\log_2 M$) bps/Hz. Its decoding decision metric in (6) can be reduced to:

$$\begin{aligned}\hat{\mathbf{U}}_t &= \arg\min_{c_1,\ldots,c_K \in \mathcal{M}} \mathrm{tr}\left[a_{t-1}^{-2}\mathbf{R}_{t-1}^{\mathrm{H}}\mathbf{R}_{t-1}\left(\sum_{i=1}^{K}\frac{\left|c_i\right|^2}{K}\mathbf{I}_{N_T}\right) - 2a_{t-1}^{-1}\,\mathrm{Re}\left(\mathbf{R}_t^{\mathrm{H}}\mathbf{R}_{t-1}\left(\sum_{i=1}^{K}c_i^{\mathrm{R}}\mathbf{A}_i + jc_i^{\mathrm{I}}\mathbf{B}_i\right)\right)\right] \\ &= \arg\min_{c_1,\ldots,c_K \in \mathcal{M}} \sum_{i=1}^{K}\mathrm{tr}\left[a_{t-1}^{-2}\mathbf{R}_{t-1}^{\mathrm{H}}\mathbf{R}_{t-1}\left(\frac{\left|c_i\right|^2}{K}\mathbf{I}_{N_T}\right) - 2a_{t-1}^{-1}\,\mathrm{Re}\left(\mathbf{R}_t^{\mathrm{H}}\mathbf{R}_{t-1}\left(c_i^{\mathrm{R}}\mathbf{A}_i + jc_i^{\mathrm{I}}\mathbf{B}_i\right)\right)\right]\end{aligned} \quad (15)$$

where the $\mathbf{A}_i$ and $\mathbf{B}_i$ are the dispersion matrices defined in (9). Due to the linear code structure of MDC-QOSTBC, the individual symbols $c_i$ in $\mathbf{U}_t$ can be separately detected using the simplified expression below:

$$\hat{c}_i = \arg\min_{c_i \in \mathcal{M}} \mathrm{tr}\left[a_{t-1}^{-2}\mathbf{R}_{t-1}^{\mathrm{H}}\mathbf{R}_{t-1}\left(\frac{\left|c_i\right|^2}{K}\mathbf{I}_{N_T}\right) - 2a_{t-1}^{-1}\,\mathrm{Re}\left(\mathbf{R}_t^{\mathrm{H}}\mathbf{R}_{t-1}\left(c_i^{\mathrm{R}}\mathbf{A}_i + jc_i^{\mathrm{I}}\mathbf{B}_i\right)\right)\right] \quad (16)$$

As shown in (16), the proposed differential MDC-QOSTBC modulation scheme is single-complex-symbol-decodable as its decoding can be achieved by the joint detection of two real symbols, $c_i^{\mathrm{R}}$ and $c_i^{\mathrm{I}}$, or one complex symbol equivalently. This results in a lower decoding complexity than many existing DSTMs [1-3, 5-6, 10], which require a larger joint detection search space.

*C. Design of Constellation Set*

The solution of the three code design criteria in (14) will be discussed in this section. As shown in Figure 1, the solutions of ($x_k$, $y_k$) to (14)(ii) can be viewed as points on $L$ concentric circles with radius $r_1$, $r_2$, ..., $r_L$, where $r_i$ are subject to the constraint:

$$\sum_{i=1}^{L} r_i^2 = L \qquad (17)$$

so as to achieve $E\left(x_k^2 + y_k^2\right) = E(r_k^2) = 1$. Next, the solution loci to (14)(i) can be represented as a hyperbola. Hence the intersection points of the hyperbola with the concentric circles represent constellation points that satisfy both (14)(i) and (ii). Every circle has at most four intersection points with the hyperbola, illustrated as $A_i$, $B_i$, $C_i$ and $D_i$, where $1 \le i \le L$, in Figure 1. However, due to their geometrical symmetry, the pair of constellation points $A_i$ and $B_i$ or $C_i$ and $D_i$ on the same circle will lead to a zero value for $\det_{\min}$ in (13) and the resultant DSTM will not have full diversity. Hence, only the pair of constellation points $A_i$ and $C_i$, or $B_k$ and $D_k$ (where $i \neq k$), are permissible on each circle, as indicated by the different markers (x and o) in Figure 1. Finally, to further comply with (14)(iii) in order to maximize the $\det_{\min}$ value, we propose to use the ($A_i$, $C_i$) constellation pairs and the ($B_k$, $D_k$) constellation pairs alternately on the adjacent circles.

The above approach enables us to reduce the variables to be optimized in (14)(iii) from $M$ pairs of ($x_k$, $y_k$) to $v$ and $r_1$, $r_2$, ..., $r_L$ where $L = M/2$, since every circle eventually gives two constellation points as the solutions to (14)(i)-(iii).

We now derive the optimum $v$ and $r_i$ values for the case of four constellation points. Referring to Figure 2, we model the points $A_1$ and $C_1$ (i.e. points A and C on the first circle) as $A_1 = r_1 \times \exp(j\theta_1)$ and $C_1 = -r_1 \times \exp(j\theta_1)$, and the points $B_2$ and $D_2$ (i.e. points B and D on the second circle) as $B_2 = r_2 \times \exp(j\theta_2)$ and $D_2 = -r_2 \times \exp(j\theta_2)$, $0 < \theta_1, \theta_2 < \pi/2$. First we focus on the optimum value for $\theta_1$ corresponding to Condition 14(iii):

$$\begin{aligned} \triangle x_{A_1C_1} &= \text{real}(A_1) - \text{real}(C_1) = 2r_1\cos(\theta_1) \\ \triangle y_{A_1C_1} &= \text{imag}(A_1) - \text{imag}(C_1) = 2r_1\sin(\theta_1) \end{aligned}$$

$$\begin{aligned}&\left[\left(\triangle x_{\mathrm{A_1C_1}}\right)^2-\left(\triangle y_{\mathrm{A_1C_1}}\right)^2\right]^{N_T}\\&=\left[\left(2r_1\cos(\theta_1)\right)^2-\left(2r_1\sin(\theta_1)\right)^2\right]^{N_T}\\&=\left[4r_1^2\left[\cos^2(\theta_1)-\sin^2(\theta_1)\right]\right]^{N_T}\\&=\left[4r_1^2\cos(2\theta_1)\right]^{N_T}\end{aligned} \tag{18}$$

To maximize (18) with respect to $\theta_1$ for even $N_T$, $\theta_1$ has to be 0 or $\pi/2$ (i.e. $A_1$ and $C_1$ lie on the x-axis). This corresponds to $v$ = 0. Next, we consider the optimum value for $\theta_2$:

$$\begin{aligned}\triangle x_{\mathrm{A_1B_2}}&=\mathrm{real}(\mathrm{A}_1)-\mathrm{real}(\mathrm{B}_2)=r_1\cos(\theta_1)-r_2\cos(\theta_2)\\\triangle y_{\mathrm{A_1B_2}}&=\mathrm{imag}(\mathrm{A}_1)-\mathrm{imag}(\mathrm{C}_2)=r_1\sin(\theta_1)-r_2\sin(\theta_2)\end{aligned}$$

$$\begin{aligned}&\left[\left(\triangle x_{\mathrm{A_1B_2}}\right)^2-\left(\triangle y_{\mathrm{A_1B_2}}\right)^2\right]^{N_T}\\&=\left[\left(r_1\cos(\theta_1)-r_2\cos(\theta_2)\right)^2-\left(r_1\sin(\theta_1)-r_2\sin(\theta_2)\right)^2\right]^{N_T}\\&=\left[r_1^2\cos(2\theta_1)-2r_1r_2\cos(\theta_1+\theta_2)+r_2^2\cos(2\theta_2)\right]^{N_T}\end{aligned} \tag{19}$$

To maximize (19) with respect to $\theta_1$ and $\theta_2$, $\theta_1+\theta_2$ has to be $\pi/2$ with the constraint that $A_1$ and $B_2$ lie in the first quadrant (as they are the intersection points of a hyperbola with the concentric circles). So, $\theta_1=0$ and $\theta_2=\pi/2$, or $\theta_1=\pi/2$ and $\theta_2=0$, are two possible solutions. Again they correspond to $v$ = 0. Without loss of generality, we choose $\theta_1=0$ and $\theta_2=\pi/2$, then equations (18) and (19) become:

$$\begin{aligned}&\text{(i)}\left[\left(\triangle x_{\mathrm{A_1C_1}}\right)^2-\left(\triangle y_{\mathrm{A_1C_1}}\right)^2\right]^{N_T}=\left[4r_1^2\right]^{N_T}\\&\text{(ii)}\left[\left(\triangle x_{\mathrm{A_1B_2}}\right)^2-\left(\triangle y_{\mathrm{A_1B_2}}\right)^2\right]^{N_T}=\left[r_1^2-r_2^2\right]^{N_T}\end{aligned} \tag{20}$$

To comply with (14)(iii), we need to maximize both terms in (20) under the constraint (17) of $\sum_{i=1}^{L} r_i^2 = L$, i.e.

$$\max\left\{\left[4r_1^2\right]^{N_T},\left[r_1^2-r_2^2\right]^{N_T}\right\}\text{ subject to } r_1^2+r_2^2=2 \tag{21}$$

The maximization in (21) can be carried out by substituting $r_2^2 = 2 - r_1^2$ and setting $\left|4r_1^2\right| = \left|r_1^2 - r_2^2\right|$. The solutions are $r_1 = \sqrt{1/3} = 0.5774$ and $r_2 = \sqrt{5/3} = 1.291$. Due to symmetry, this derivation applies to $A_1$ and $D_2$, or $C_1$ and $B_2$, too.

For verification purpose, we have carried out numerical optimizations on computer to obtain the coding gain of the proposed DSTM scheme for different values of *v*. The numerical solutions obtained for *M* = 4 and 8 constellation points are shown in Figure 3. They show that the coding gain increases monotonically as *v* decreases, and *v* = 0 gives the best coding gain for both values of *M*. This validates the above derivation for the case of *M* = 4 constellation points. For the case of *M* = 8 constellation points, analytical derivation becomes much more tedious, hence we rely on the numerical optimization results shown in Figure 3.

It should be pointed out that (14)(iii) is a max-min optimization problem, whose numerical solutions depend on the initial values and may give only local maxima. To ensure that the solutions converge, we have used several different initial values in our optimization. Hence we believe that the coding gain values shown in Figure 3 are close to the global maxima. In Figure 4, we show the two optimized constellation sets $\mathcal{M}_1$ and $\mathcal{M}_2$ with *M* = 4 and 8 constellation points respectively. Note that these constellation points lie on the x- or y-axis due to *v* = 0, but they are not the standard regular QPSK or Amplitude-PSK constellations. Furthermore, in contrast to coherent MDC-QOSTBC which requires constellation rotation to achieve full diversity [11, 12, 13], the DSTM constellations shown in Figure 4 already achieve full diversity by design, hence they are to be used *without additional constellation rotation*.

## IV. PERFORMANCE

In our simulations, the channel is assumed to be flat fading and quasi-static. To facilitate result comparison with [10], every frame from each antenna is assumed to have 132 symbols.

In Figure 5, we compare the block error rate (BLER) performances of our proposed single-symbol-decodable DSTM scheme and the DSTM scheme based on O-STBC [8][9] under common spectral efficiency values. For the case of eight transmit antennas and spectral efficiency of 1.5 bps/Hz, our proposed DSTM scheme based on rate-3/4 MDC-QOSTBC and four-point constellation set $\mathcal{M}_1$ outperforms the DSTM scheme based on rate-1/2 O-STBC and 8QAM constellation, with a 1.5 dB gain at BLER = $10^{-4}$. For the case of four transmit antennas and spectral efficiency of 2 bps/Hz, our proposed DSTM scheme based on rate-1 MDC-QOSTBC and four-point constellation set $\mathcal{M}_1$ again outperforms the DSTM scheme based on rate-1/2 O-STBC and 16QAM constellation, with a 3 dB gain at BLER = $10^{-4}$. These results show that our proposed DSTM scheme is able to offer good performance gain over DSTM schemes based on rate-1/2 O-STBC. For four transmit antennas and spectral efficiency of 3 bps/Hz, our proposed DSTM scheme with rate-1 MDC-QOSTBC and eight-point constellation set $\mathcal{M}_2$ performs comparably with the DSTM scheme based on rate-3/4 O-STBC and 16QAM constellation. This is because QAM constellation has better Euclidean distances than the "Amplitude-PSK-like" constellation of $\mathcal{M}_2$ when the constellation size is large. Hence the performance gain of our proposed DSTM over differential O-STBC diminishes at high spectral efficiency levels, a phenomenon encountered by the DSTM scheme reported in [10] too. Finally, we have also included in Figure 5 the "Genie" BLER results which are obtained by assuming that the exact values of $a_{t-1}$ are available to the DSTM decoding process in (16). These "Genie" results show that the effects of error propagation due to the estimation of $a_{t-1}$ in (6) are negligible, hence we will use the practical (non-Genie) DSTM decoder in the subsequent study.

In Figure 6, we compare the BLER performance and *decoding complexity* of our proposed single-symbol-decodable DSTM scheme with the DSTM schemes based on rate-1 QO-STBC [5][10] and Sp(2) [3], for four transmit and one receive antennas. First, we compare the BLER performance of the DSTM scheme from [5][10] with our DSTM scheme. Under a spectral efficiency of 3 bps/Hz, our DSTM scheme performs comparably as the DSTM scheme from [10], but the decoding search space of the latter is the square of ours (i.e. our DSTM decoder needs to search over 8 constellation points while the DSTM decoder of [10] needs to search over all combinations of two sets of 8 constellation points, which amounts to a search space of 64). Likewise can be said for the case of the DSTM scheme from [5] with spectral efficiency of 2 bps/Hz. Next, when compared against the Sp(2) DSTM scheme from [3], our DSTM scheme with spectral efficiency of 3 bps/Hz has a 0.5dB performance gain at BLER $10^{-4}$ over the Sp(2) DSTM scheme with spectral efficiency of 3.13 bps/Hz. Hence both schemes can be considered to have comparable decoding performance, but our DSTM scheme is single-symbol decodable with a decoding search space of 8, while the Sp(2) DSTM scheme has a much larger decoding search space of 5929. At a lower spectral efficiency, our DSTM scheme does not perform as good as the Sp(2) DSTM scheme with 1.95 bps/Hz and a decoding search space of 225, but our DSTM scheme has a slightly higher spectral efficiency of 2 bps/Hz and significantly smaller decoding search space of 4. These results suggest that our single-symbol-decodable DSTM scheme is able to offer significant reduction in decoding complexity with comparable decoding error probability over the DSTM schemes in [3, 5, 10] at similar spectral efficiency.

## V. CONCLUSIONS

We have shown that a single-symbol-decodable differential space-time modulation (DSTM) scheme with full diversity can be constructed using the Minimum-Decoding-Complexity Quasi-Orthogonal STBC (MDC-QOSTBC). The main idea is to force the MDC-QOSTBC codeword to be quasi-unitary by customizing its symbol constellation. We have derived the constellation design criteria and obtained optimized solutions for the corresponding constellation set. Our proposed DSTM scheme has the merits of MDC-QOSTBC, hence it achieves higher code rate (rate 1 for four transmit and rate 3/4 for eight transmit antennas) than a DSTM scheme based on O-STBC, and much lower decoding complexity than DSTM schemes based on Sp(2) and double-symbol-decodable QO-STBC. The code rate advantage leads to a better error probability performance than DSTM based on rate-1/2 O-STBC, while the decoding complexity advantage is achieved with no or little trade-off in decoding error probability performance.

## Acknowledgment

The authors would like to thank the anonymous reviewers for their comments that greatly improve this paper.

# Figures

*List of figures:*

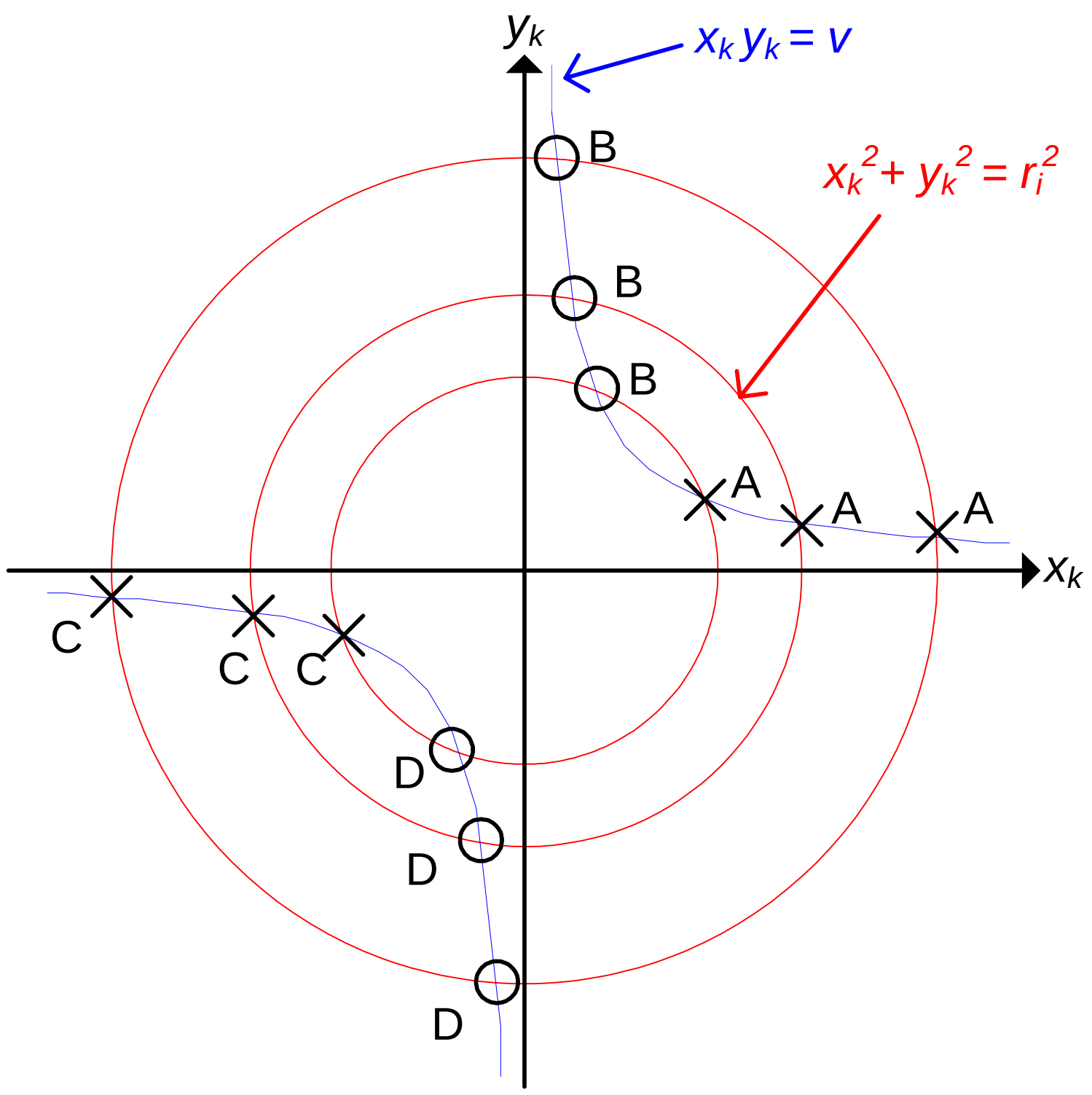

Figure 1 Solution loci of the DSTM constellation design criteria (14)(i) and (ii)

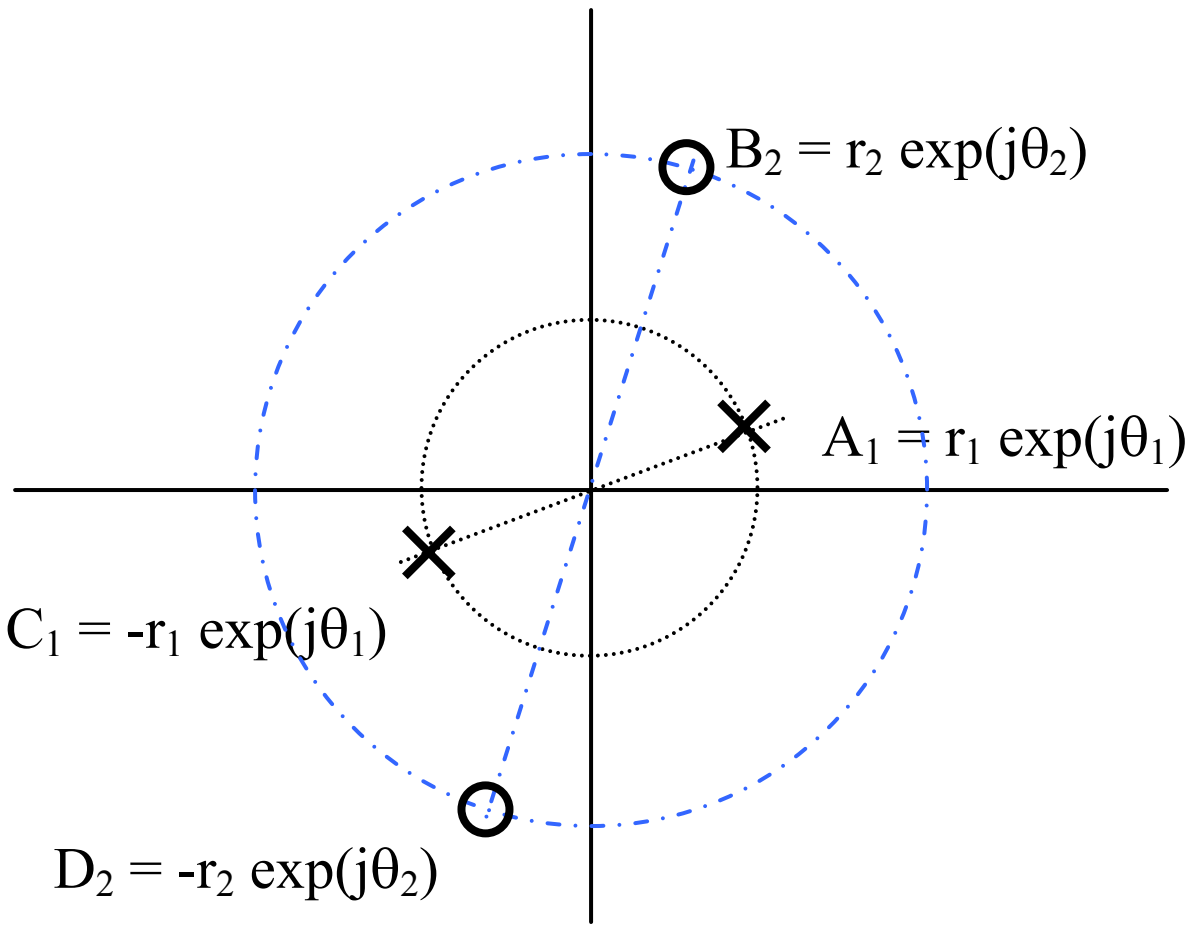

Figure 2 Pre-optimized constellation points for the proposed DSTM

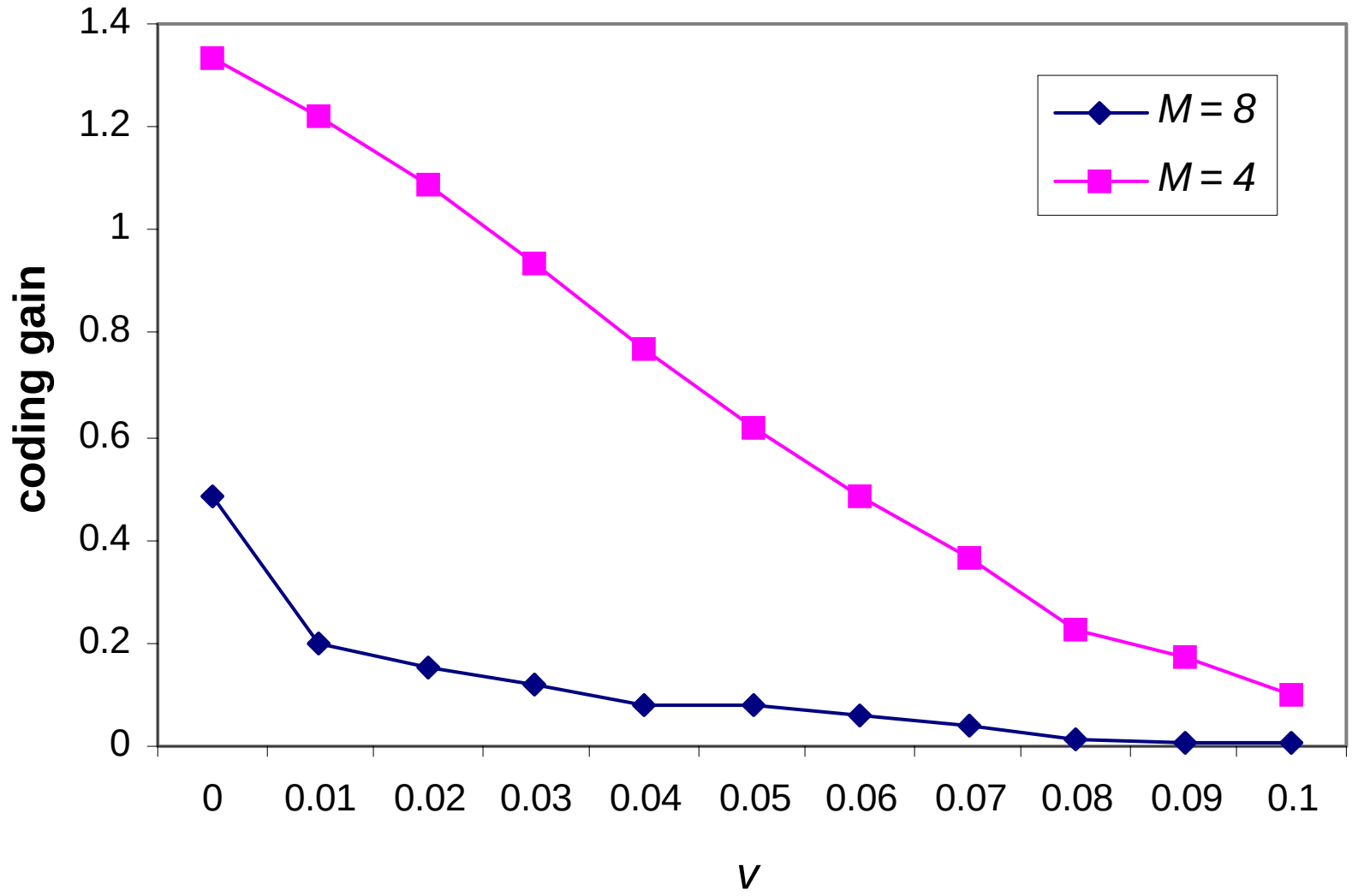

Figure 3 Optimization of coding gain through numerical optimization

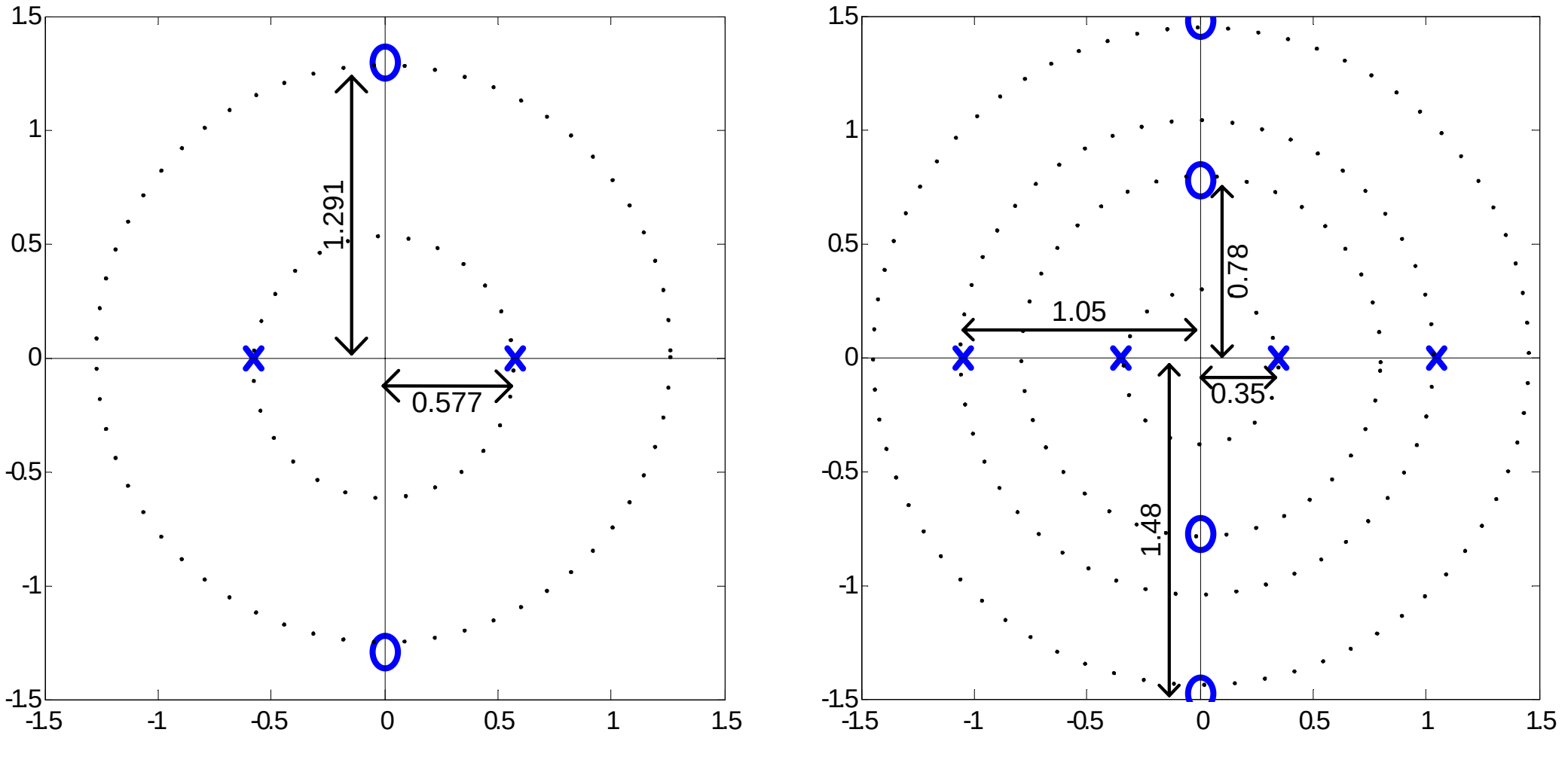

(a) $\mathcal{M}_1$ with $M = 4$ constellation points (b) $\mathcal{M}_2$ with $M = 8$ constellation points

Figure 4 Optimized constellation sets for DSTM with MDC-QOSTBC

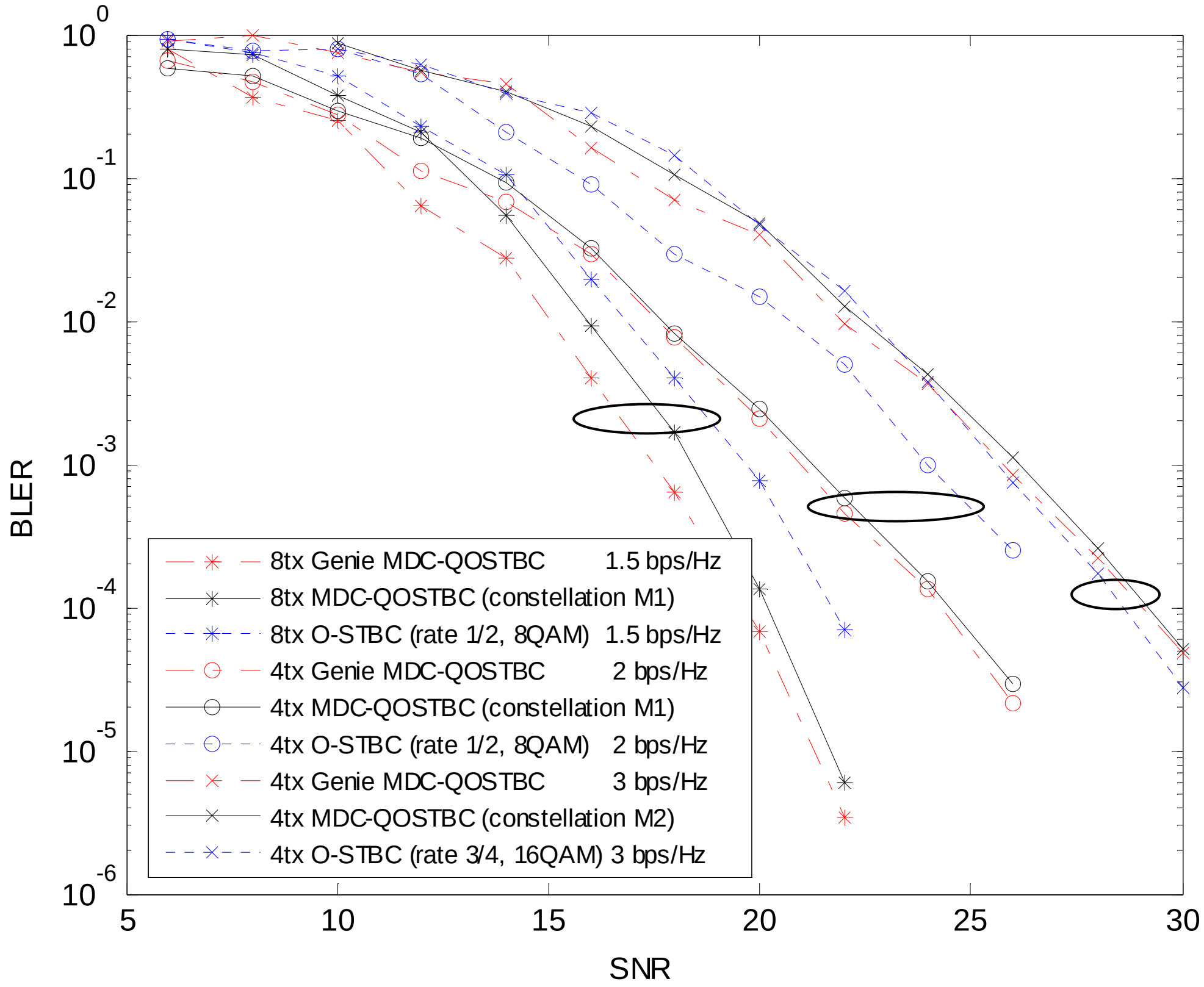

Figure 5 Block error rate (BLER) of DSTM schemes based on O-STBC and MDC-QOSTBC

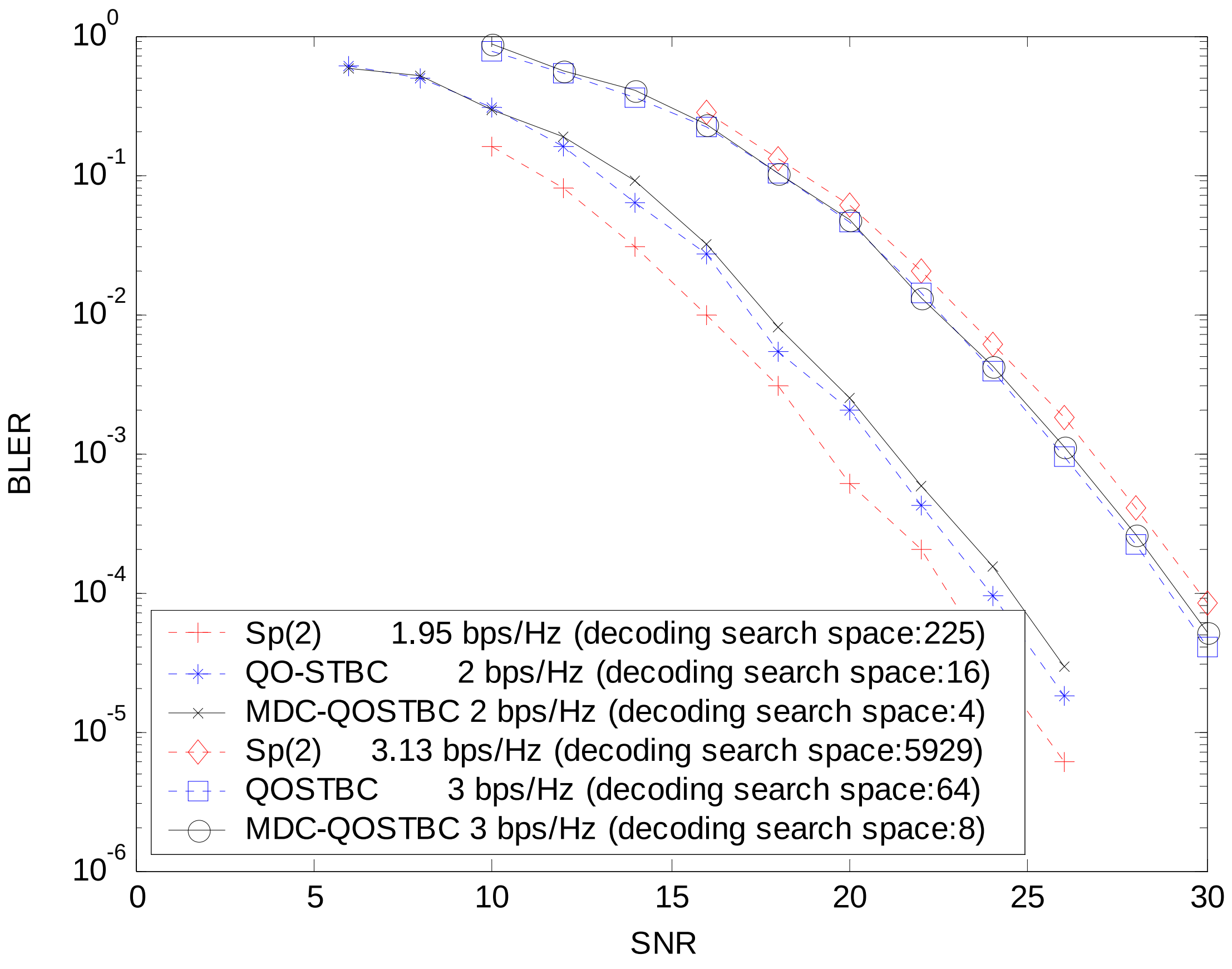

Figure 6 Block error rate (BLER) of DSTM schemes for four transmit and one receive antennas